\documentclass[authoryear,12pt]{elsarticle}
\usepackage{graphicx}
\begin{document}

\newcommand{\ii}{\'{\i}}

\def\lsim{\ ^{<}\!\!\!\!_{\sim}\>}
\def\gsim{\ ^{>}\!\!\!\!_{\sim}\>}

\centerline{\Large \bf Assessing the physical nature of near-Earth asteroids}

\centerline{\Large \bf through their dynamical histories} 

\vspace{1cm}

\centerline{Julio A. Fern\'andez$^*$, Andrea Sosa, Tabar\'e Gallardo, Jorge N. Guti\'errez}

\bigskip
\bigskip

\centerline{Departamento de Astronom\ii a, Facultad de Ciencias,}

\centerline{Universidad de la Rep\'ublica, Igu\'a 4225, 14000 Montevideo, Uruguay}

\vspace{2cm} 

\centerline{Accepted for publication in ICARUS}
\vspace{0.5cm}
\centerline{April 2014}

\vspace{5cm}

\noindent Number of pages: 28\\
Number of figures: 13\\
Number of tables: 3

\vspace{5cm}

\noindent $^*$ Corresponding author.\\
E-mail address: julio@fisica.edu.uy

\vfill
\eject

\centerline{\bf Abstract}

\bigskip

\noindent We analyze a sample of 139 near-Earth asteroids (NEAs), defined as those that reach perihelion distances $q < 1.3$ au, and that also fulfill the conditions of approaching or crossing Jupiter's orbit (aphelion distances $Q > 4.8$ au), having Tisserand parameters $2 < T < 3$  and orbital periods $P < 20$ yr. In order to compare the dynamics, we also analyze a sample of 42 Jupiter family comets (JFCs) in near-Earth orbits, i.e. with $q < 1.3$ au. We integrated the orbits of these two samples for $10^4$ yr in the past and in the future. We find that the great majority of the NEAs move on stable orbits during the considered period, and that a large proportion of them are in one of the main mean motion resonances with Jupiter, in particular the 2:1. We find a strong coupling between the perihelion distance and the inclination in the motion of most NEAs, due to Kozai mechanism, that generates many sungrazers. On the other hand, most JFCs are found to move on very unstable orbits, showing large variations in their perihelion distances in the last few $10^2 - 10^3$ yr, which suggests a rather recent capture in their current near-Earth orbits. Even though most NEAs of our sample move in typical 'asteroidal' orbits, we detect a small group of NEAs whose orbits are highly unstable, resembling those of the JFCs. These are: 1997 SE5, 2000 DN1, 2001 XQ, 2002 GJ8, 2002 RN38, 2003 CC11, 2003 WY25, 2009 CR2, and 2011 OL51. These objects might be inactive comets, and indeed 2003 WY25 has been associated with comet Blanpain, and it is now designed as comet 289P/Blanpain. Under the assumption that these objects are inactive comets, we can set an upper limit of $\sim 0.17$ to the fraction of NEAs with $Q > 4.8$ au of cometary origin, but it could be even lower if the NEAs in unstable orbits listed before turn out to be {\it bona fide} asteroids from the main belt. This study strengthens the idea that NEAs and comets essentially are two distinct populations, and that periods of dormancy in comets must be rare. Most likely, active comets in near-Earth orbits go through a continuous erosion process in successive perihelion passages until disintegration into meteoritic dust and fragments of different sizes. In this scenario, 289P/Blanpain might be a near-devolatized fragment from a by now disintegrated parent comet.

\bigskip


\vfill
\eject

\section{Introduction}

The traditional difference between comets and asteroids based on the type of orbits (e.g. the Tisserand parameter) and/or on whether they show or not gaseous and/or dust activity has become increasingly blurry with the unexpected discovery of activity on some typical main-belt asteroids, the so-called main-belt comets \citep{Hsie06}, and the discovery of objects on typical cometary orbits that do not show any activity at all. The spectra of some bodies of the latter group with low values of the Tisserand parameter $T$ ($<2.7$), typical of Jupiter family comets, are found to be very red, compatible with dead or dormant comets, but also with Trojan and Hilda asteroids \citep{Lica08}.\\ 

The lack of observable activity is particularly striking in the case of bodies that approach the Earth, the Near-Earth Asteroids (NEAs), for which one should expect to find activity if they contain volatiles, namely if they are of cometary nature, once they are exposed to the more intense Sun's radiation. One way out of the puzzle is to assume that comets build insulating dust mantles after their perihelion passages, that turn them into inactive, asteroid-looking bodies \citep{Shul72, Brin80, Rick90}. Therefore, the possibility that comets pass through periods of dormancy, or become extinct is one of the issues to solve, and also how widespread is the phenomenon. This has to be confronted with the observation of several comets that disintegrated near  perihelion owing to their volatile composition and fragile structure \citep{Seka84, Weav01, Batt13}. Several authors have suggested a possible cometary origin for some or most NEAs \citep{Weth88, Levi94, Deme08}, though it has been argued that the transfer process of bodies from the asteroid belt to NEA orbits is efficient enough to keep the current observed population of NEAs in steady-state without needing to invoke an extra comet source \citep{Rabi97, Fern02}. The similarities between the spin rate and shape distribution of NEAs and main-belt asteroids also led \citet{Binz92} to conclude that most NEAs must come from the main belt.\\

We have two possible scenarios for the physical evolution of periodic comets in which: 1) they get insulating dust mantles during their perihelion passages becoming dormant or extinct; or 2) they keep active all the way until disintegration into meteoritic dust and, perhaps, leaving behind large fragments of devolatized material (Fig. 1). The first scenario may also foresee the possibility of intermittent activity, during which the comet passes through alternate periods of dormancy and activity until complete disintegration (or dynamical ejection, or collision with the Sun or a planet).\\

\begin{figure}[h]
\resizebox{9cm}{!}{\includegraphics{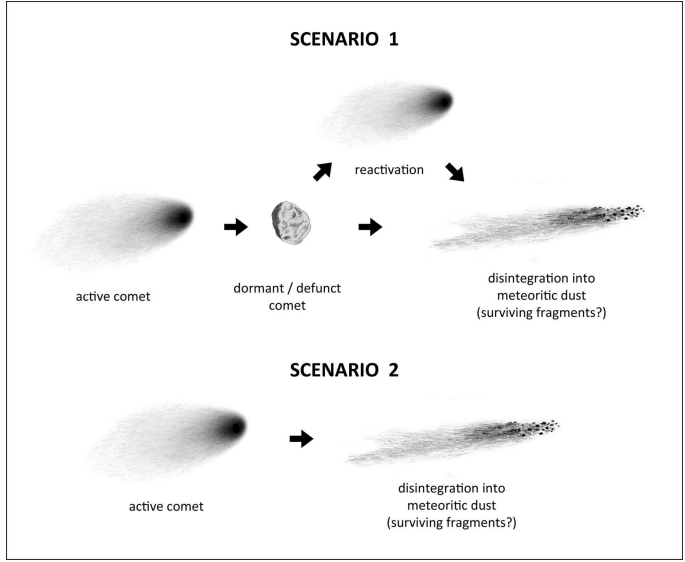}}
\caption{Two possible scenarios for the physical evolution of comet nuclei in the inner planetary region. In both cases, and unless the comet is ejected or collides with the Sun or a planet first, disintegration will be the ultimate fate after a varying number of perihelion passages.}
\label{}
\end{figure}

We plan to deal with bodies of different characteristics (asteroids, comets, Earth-approaching objects, main-belt objects, etc.) so, in order to simplify the language, we will use different acronyms to refer to them. A summary of the acronyms used here is presented in Table I. We note that the acronym 'NEA' might include objects of cometary origin that happen to be inactive at present.\\


\begin{table}[htb]
\centerline{Table I: Glosary of acronyms used in the text}
\begin{center}
\begin{tabular}{l l} \hline
Acronym & Description \\ \hline
NEO & Near-Earth Object ($q < 1.3$ au) that involves bodies of different nature\\ 
    & (comets, asteroids, meteoroids).\\
NEA & Near-Earth Asteroid ($q < 1.3$ au), usually referred to bodies that look\\    & inactive. \\
JFC & Jupiter Family Comet \\
NEJFC & Near-Earth Jupiter Family Comet ($q < 1.3$ au) \\
NEACO & NEA in Cometary Orbit \\ \hline
\end{tabular}
\end{center}
\end{table}

This work has been motivated by a previous work \citep{Sosa12} that studied the time-evolution of the average perihelion distance, $<q>$, of samples of JFCs with $q<1.3$ au and NEAs for $10^3$ yr in the past and in the future. A comparison of the time evolution of $<q>$ between NEAs and JFCs showed striking differences: it stayed more or less constant during the studied period for NEAs, whereas it showed a steep increase in the past and a more moderate increase in the future for JFCs. This  asymmetry in the evolution of $<q>$ for JFCs was interpreted as due to the short physical lifetime of JFCs with $q < 2$ au, of a few $10^3$ yr to about $10^4$ yr, that favors the discovery of those comets that evolve fast enough to low-$q$ orbits (where they are more easily detected) before they disintegrate. Our aim in this work is to extend the integrations for a longer period ($\pm 10^4$ yr) and to check if the NEAs have indeed dynamical evolutions quite different from those showed by JFCs. A byproduct of this study is to detect potential inactive comets among NEAs.  

\section{The method}

\subsection{The samples and data sets}

We analyzed a sample of 139 NEAs with the same orbital characteristics as the JFCs, namely Tisserand parameters $2 < T < 3$, and orbital periods $P < 20$ yr. Furthermore, we imposed the condition that they are Jupiter-approaching or crossing objects, namely with aphelion distances $Q > 4.8$ au. We also restrict the sample to those orbits of better quality, as given by the condition codes $\leq 5$ in the JPL scale 0-9 (from the best to the poorest quality). For comparison purposes, we also studied a sample of 42 NEJFCs in orbits with the same constraints as those for the asteroid sample. The orbits of NEJFCs were integrated neglecting nongravitational (NG) forces. To check the influence of these forces on the evolution, we also integrated the orbits including NG terms in the cases they were estimated. We did not find significant differences, in statistical terms, in the orbital evolution of comets with and without NG forces. A more thorough discussion of this topic will be given in a forthcoming paper. The orbital data were extracted from the NASA/JPL Small-Body Database\footnote{http://ssd.jpl.nasa.gov/sbdb.cgi}, as known by the end of 2012.

\subsection{The numerical integrations}

We integrated the orbits of the considered objects in a heliocentric frame for $10^4$ yr, in the past and in the future with respect to the present epoch, which was defined as JD 2456200.500, i.e. CE 2012 September 30, 00:00:00 UT, Sunday. The output interval was 1 yr. We considered for each object five clones, where each clone was generated by means of a random Gaussian distribution in the  6-orbital parameters space, with a mean value equal to the nominal osculating value for the present epoch, and a standard deviation $\sigma$ equal to the nominal uncertainty. If the object and/or any of the clones showed large changes in the perihelion distance and semimajor axis $a$, we then integrated other 50 clones of the object in order to analyze its dynamical behavior more in detail, in particular to check how sensitive the changes in $q$ and $a$ are to small variations in the initial conditions.\\

The integrations were performed with the orbital integrator MERCURY \citep{Cham99}, using the general Bulirsch-Stoer code as the N-body algorithm, which has been proven to be accurate in most situations, especially when very close encounters with a planet are involved. All the minor bodies were considered as point masses. Only Newtonian gravitational forces (including those from the eight planets) were considered. The sample objects were considered as 'ejected' if they reached a heliocentric distance of 100 au. We considered that a sample body had a close encounter with a given planet when it reached a planetocentric distance smaller than 3 Hill radii, and each time that this condition was fulfilled, the encounter conditions were stored for further analysis. 

\subsection{Search and analysis of resonances}

In order to identify possible resonant motions we proceeded as follows. When the semimajor axis oscillated around a quasi-constant value $a_0$, we calculated the strengths for all possible resonances located near $a_0$ following \citet{Gall06}. Then, we computed the time evolution of the corresponding critical angles, $\sigma$, for the strongest resonances in a certain small interval around $a_0$. We recall that the critical angle for a $k$-order resonance, $|p+k|:|p|$, is given by

\begin{equation}
\sigma = (p+k)\lambda_P - p\lambda - k\varpi ,
\end{equation}
where $p$ and $k$ are integers, $\lambda_P$ and $\lambda$ are the mean longitudes of the planet and the body, respectively, and $\varpi$ is the longitude of perihelion of the body. From the previous analysis we obtained the libration center and the semiamplitude $A$ in degrees.\\

\section{The results}

\subsection{General features of the orbit evolution}

Most NEAs show an orbital evolution stable in the past $10^4$ yr. The resilience of such bodies in short-period, small-$q$ orbits over long time scales strongly suggests a rocky composition, able to withstand the intense heat for at least several thousands of passages near the Sun. Their orbits are characterized as having the perihelion distances confined within $q < 2.5$ au, and semimajor axes $a < 7.37$ au (orbital periods $P < 20$ yr) during the past $10^4$ yr (in most cases the same situation continues in the future $10^4$ yr). We will define this type of orbits as 'asteroidal'. We show in Fig. 2 a couple of NEAs representative of our sample. We plot the time evolution of the orbital elements and the distances of approach to Jupiter smaller than 3 Hill radius. On the left panels we show object 2004 RU164 that evolves within the 2:1 mean motion resonance (MMR) with Jupiter. We can see that the object avoids encounters with Jupiter at distances smaller than $\sim 1.02$ au. A weak coupling between $q$ and $i$ is observed due to the Kozai mechanism (see Section 3.4 below). At the right panels we show object 2004 QU24 evolving within a secular (nonresonant) motion with a semimajor axis confined within the range $\sim 3.32 - 3.38$ au. There is in this case a strong coupling between $q$ and $i$ due the the Kozai mechanism. The object avoids encounters with Jupiter at distances smaller than about 1 au during the studied period.\\ 

We found very often that the orbital evolution was driven by a strong resonance, though in some cases we observed temporary captures in different resonances, and there were also cases where the dynamical evolution was found to be mainly nonresonant. For all the studied objects the dominant resonances were with Jupiter. In Fig. 3 we plot the semimajor axes of NEAs on stable orbits averaged over intervals of $10^3$ yr. As mentioned before, most semimajor axes librate around values close to some of the strongest MMRs with Jupiter, specially the 2:1 MMR ($a=3.28$ au), as is the case of 2004 RU164 (Fig. 2, left). There is also some important contribution from other MMRs, like the 5:2, 7:3, 9:4 and 4:3, that stand up, after the 2:1, as the strongest ones within the interval $2.5 \lsim a \lsim 4.5$ au \citep{Gall06}. On the other hand, we note a lack of NEAs at or close to the 3:2 MMR with Jupiter ($a \simeq 3.97$ au) that might have as progenitors the Hilda asteroids. An explanation to this striking feature was provided by \citet{Disi05} who showed that Hilda asteroids escaping the resonance mainly stay in the region of perihelion distances $q > 2.5$ au, 8\% of them end up colliding with Jupiter, and the rest are ejected by the Jovian planets. Only about 6\% stay at any time as JFCs with $q < 2.5$ au.\\

By contrast to the dynamical behavior of most NEAs, most JFCs show unstable orbits with large and erratic variations of their orbital parameters, reflecting the effect of strong perturbations by Jupiter. Their residence times in small-$q$ orbits previous to their discovery are found to be short: $\lsim 10^3$ yr, consistent with physical lifetimes of $\sim 2500$ yr \citep{Fern02}. We will define this type of orbits as 'cometary'. In Fig. 4 (left panels) we show a typical case of a JFC.\\

\begin{figure}[h]
\begin{center}
\begin{tabular}{c c}
\resizebox{6cm}{!}{\includegraphics{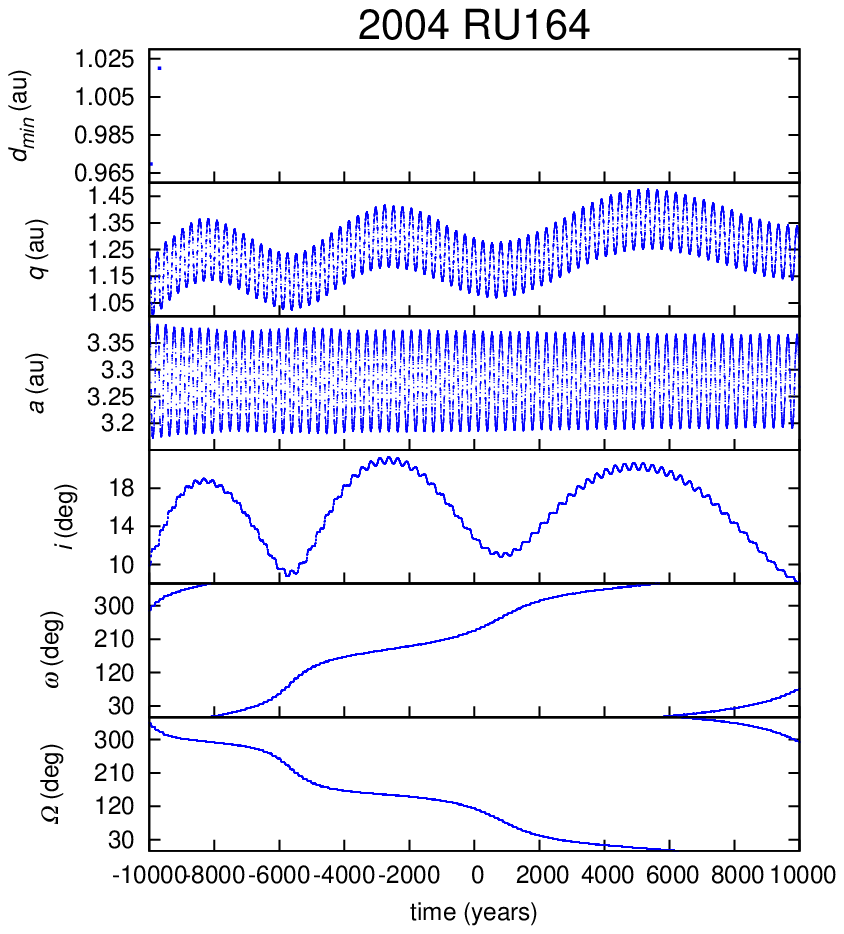}}
\label{} &
\resizebox{6cm}{!}{\includegraphics{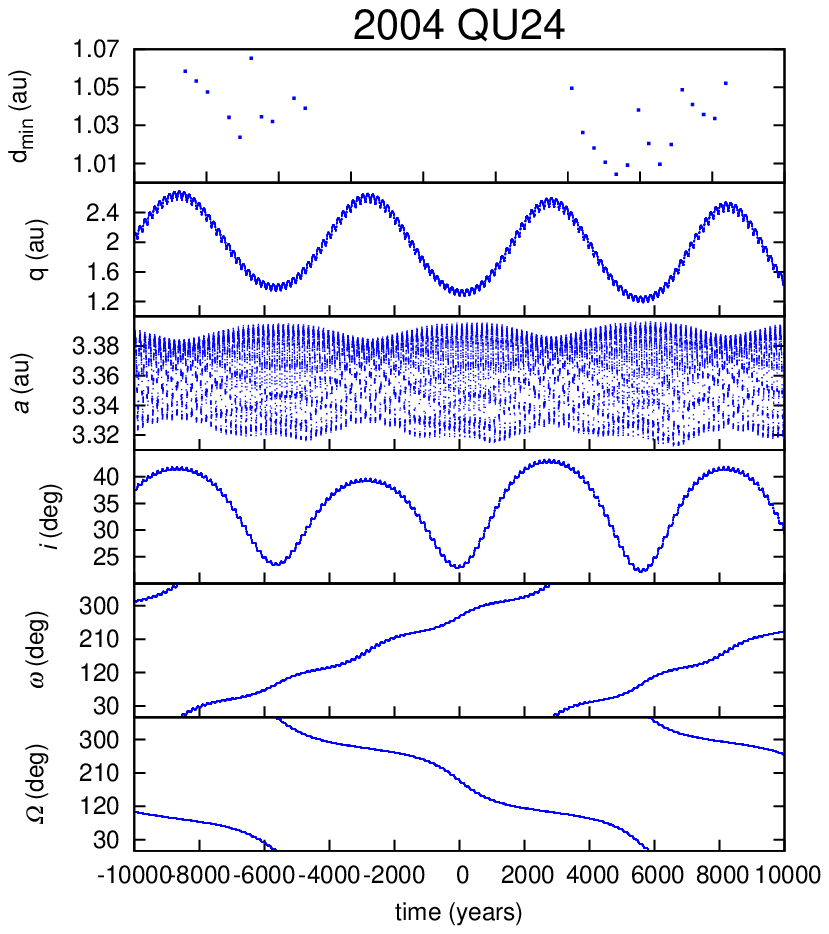}}
\label{}
\end{tabular}
\end{center}
\caption{Examples of NEAs in stable orbits.}
\end{figure}

A few NEAs show strong and erratic variations in their orbital elements, in particular in $q$ and $a$, resembling those of JFCs, so their orbits may be classified as 'cometary'. The NEA 1997 SE5 is one of these cases (Fig. 4, right panels). We can see the similarity of this orbit with that of 67P/Churyumov-Gerasimenko (Fig. 4, left panels), and, at the same time, the big difference with the NEAs on stable orbits of Fig. 2.\\

\begin{figure}[h]
\resizebox{10cm}{!}{\includegraphics{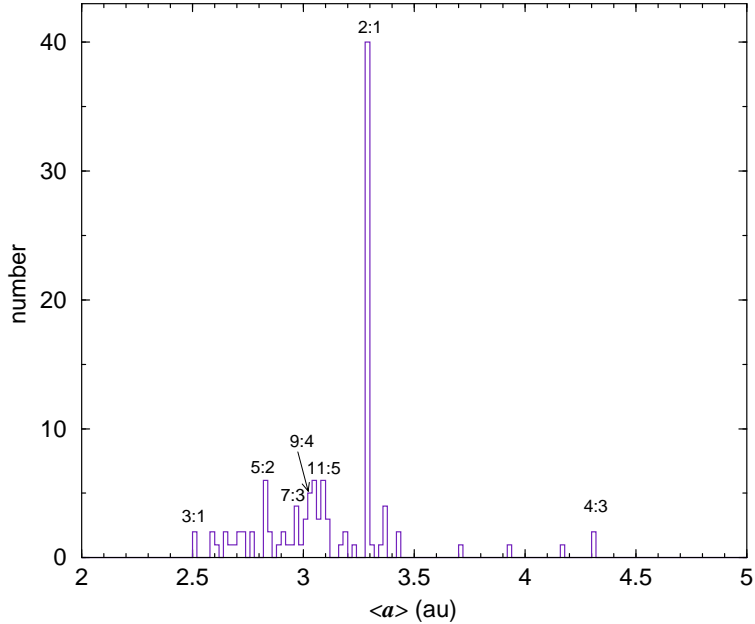}}
\caption{Distribution of average semimajor axes $<a>$ (computed within intervals of $10^3$ yr) for the sample of NEAs with $Q > 4.8$ au moving on stable orbits and their clones. We indicate the positions of the most relevant mean motion resonances with Jupiter.}
\label{}
\end{figure}

\begin{figure}[h]
\begin{center}
\begin{tabular}{c c}
\resizebox{6cm}{!}{\includegraphics{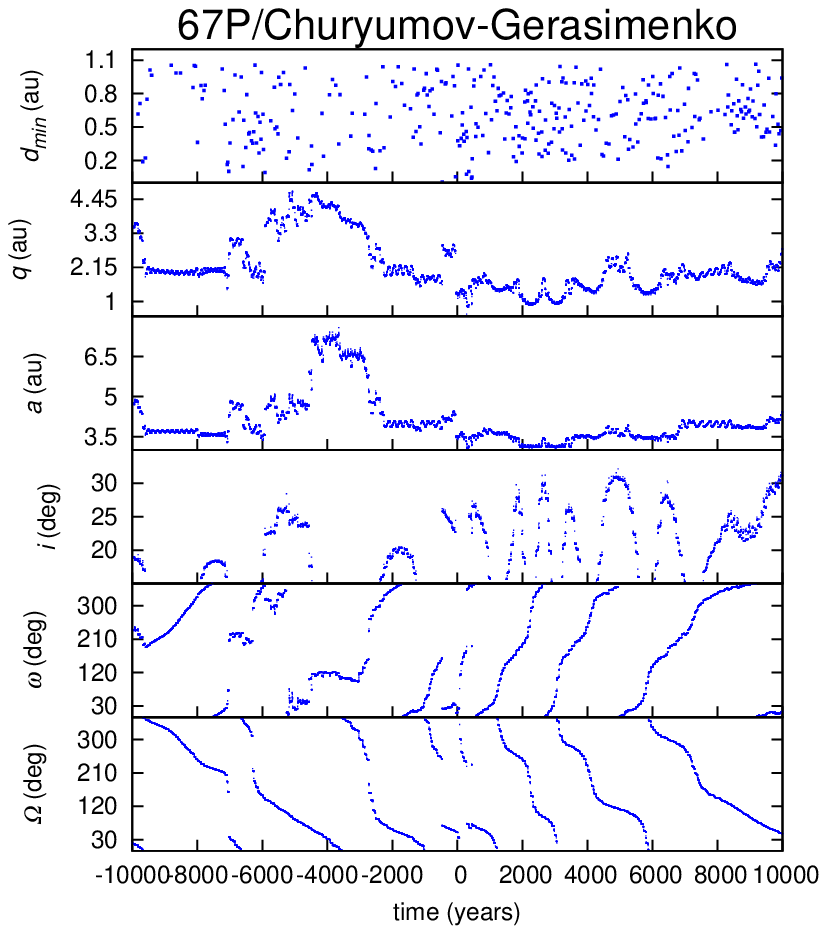}}
\label{} &
\resizebox{6cm}{!}{\includegraphics{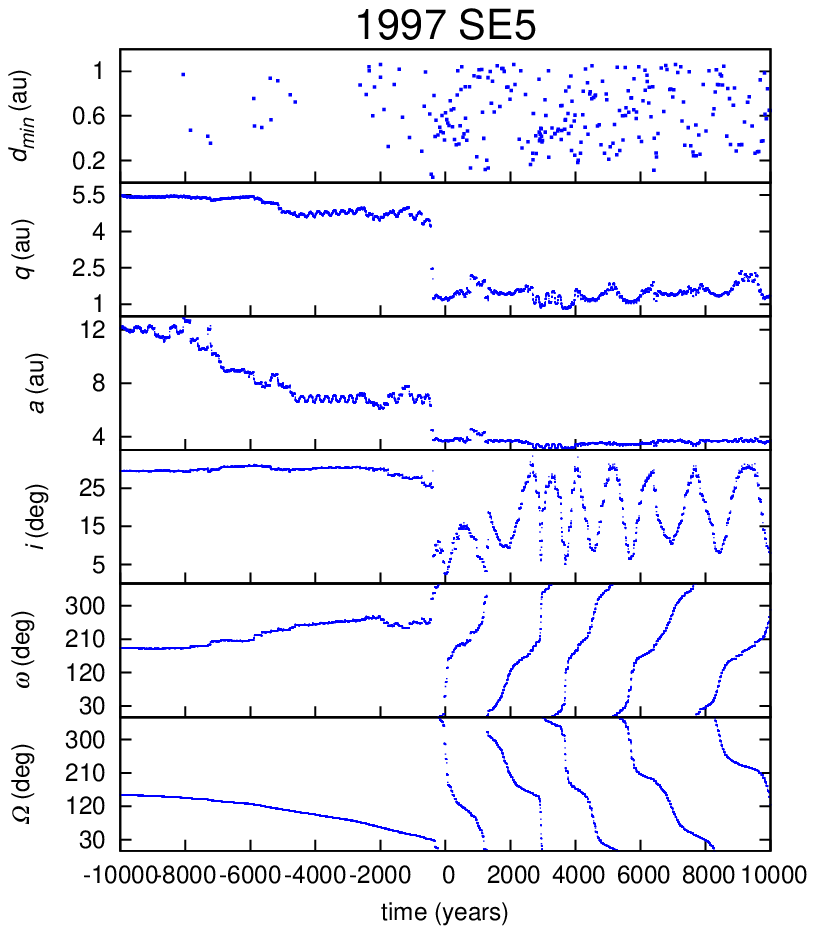}}
\label{}
\end{tabular}
\end{center}
\caption{Comet 67P/Churyumov-Gerasimenko moves on a typical unstable 'cometary' orbit (left). A few NEAs of our sample, like 1997 SE5 (right), show unstable orbits similar to those of JFCs.}
\end{figure}
 
\subsection{Characterization of the degree of variability of an orbit: Definition of the indices $f_q$ and $f_a$}

To characterize quantitatively how stable or unstable are the orbits of our sample objects, we define the index $f_q$ as the fraction of time that a given object or any of its clones spend with a perihelion distance $q > 2.5$ au, or moves on a long-period orbit that takes the object to heliocentric distances $r > 100$ au, in the past $10^4$ yr, namely

\begin{equation}
f_q = \frac{\sum_{j=1}^N \Delta t_j}{N \times 10^4},
\end{equation}
where $\Delta t_j$ is the length of time (in yr) that the object and its clones, $j=1,....,N$, have $q > 2.5$ au, or move on an orbit that reaches $r > 100$ au, in the past $10^4$ yr. For the cases of objects in stable orbits, we have $N=6$ (object + 5 clones), whereas for objects moving in unstable orbits $N=51$ (object + 50 clones) (cf. Section 2.2).\\

Similarly, we define the index $f_a$ as the fraction of time that a given object and its 50 clones spend with $a > 7.37$ au ($P > 20$ yr) in the past $10^4$ yr, i.e.

\begin{equation}
f_a = \frac{\sum_{j=1}^N \Delta t'_j}{N \times 10^4},
\end{equation}
where $\Delta t'_j$  is the length of time (in yr) that the object and its clones, $j=1,....,N$, have $a > 7.37$ au.\\ 

Obviously, objects in asteroidal orbits have $f_q = f_a = 0$, except for a few cases, to be analyzed below, in which $q$ shows periodic oscilations leading to maxima above 2.5 au (nevertheless, these anomalous cases fulfill the other condition for stable orbit $f_a = 0$).\\

\begin{figure}[h]
\resizebox{10cm}{!}{\includegraphics{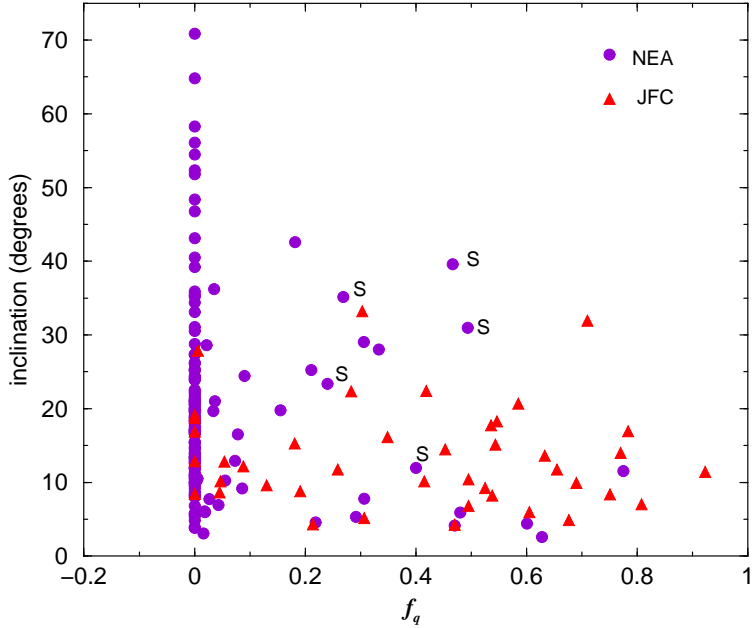}}
\caption{The fraction of time that the object remains with $q > 2.5$ au versus the inclination for NEAs and JFCs. The letter "S" stands for those NEAs in stable orbits.}
\label{}
\end{figure}

In Fig. 5 we show a plot of the inclination at $t=0$ versus the fraction $f_q$ for the samples of NEAs and JFCs. We can see that most NEAs fall in the $f_q=0$ vertical axis, with inclinations that spread over a wide range, $0^{\circ}-70^{\circ}$. On the other hand, JFCs spread over all the range $0 < f_q < 1$, with a more flattened distribution of inclinations. The letter "S" stands for the five bodies of our NEA sample, 3552 Don Quixote, 1982 YA, 2004 QU24, 2009 AT, and 2010 NW1, whose values $f_q > 0$ are misleading, since they actually move on stable orbits with large oscillations of $q$ that take them above 2.5 au. The stability of their orbits is reflected in that $f_a=0$ (for 3552 Don Quixote we got a value slightly different from zero, $f_a=0.04$).\\ 

We also wanted to investigate what is the maximum inclination $i_{max}$ that a body can reach over the studied period, and what is the relative change in the inclination $(i_{max}-i_o)/i_{max}$, where $i_o$ is the inclination at present. We plot in Fig. 6 $(i_{max}-i_o)/i_{max}$ versus $i_{max}$ for NEAs and JFCs. We see that while NEAs tend to spread more or less evenly over all the frame, JFCs tend to concentrate at the upper left corner of the diagram, which suggests that JFCs tend to be discovered when they experience strong decreases in $i$ that favor close encounters with Jupiter. On the other hand, some NEAs can attain large inclinations by the Kozai mechanism (see Section 3.4).\\

\begin{figure}[h]
\resizebox{10cm}{!}{\includegraphics{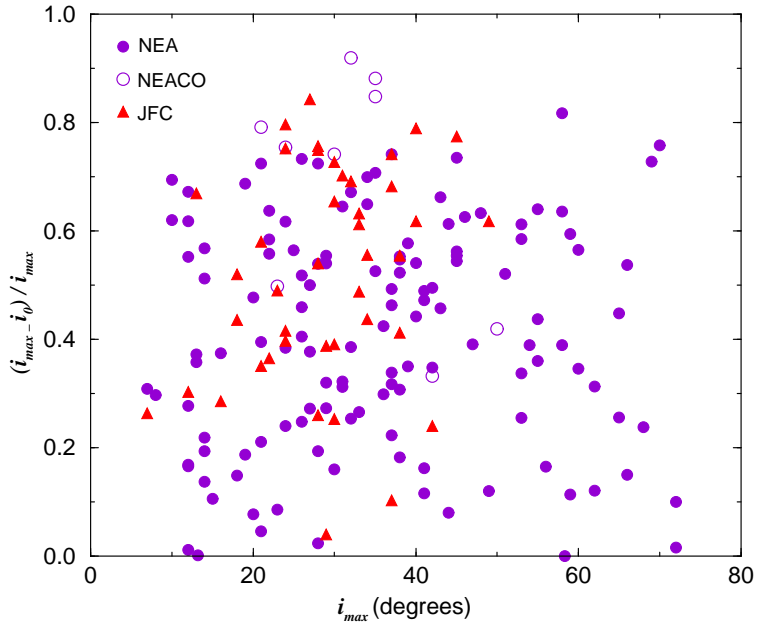}}
\caption{The maximum inclination that a body reaches during the studied period versus the relative change in the inclination (maximum - present) for NEAs, JFCs and NEACOs.}
\label{}
\end{figure}

\begin{figure}[h]
\resizebox{10cm}{!}{\includegraphics{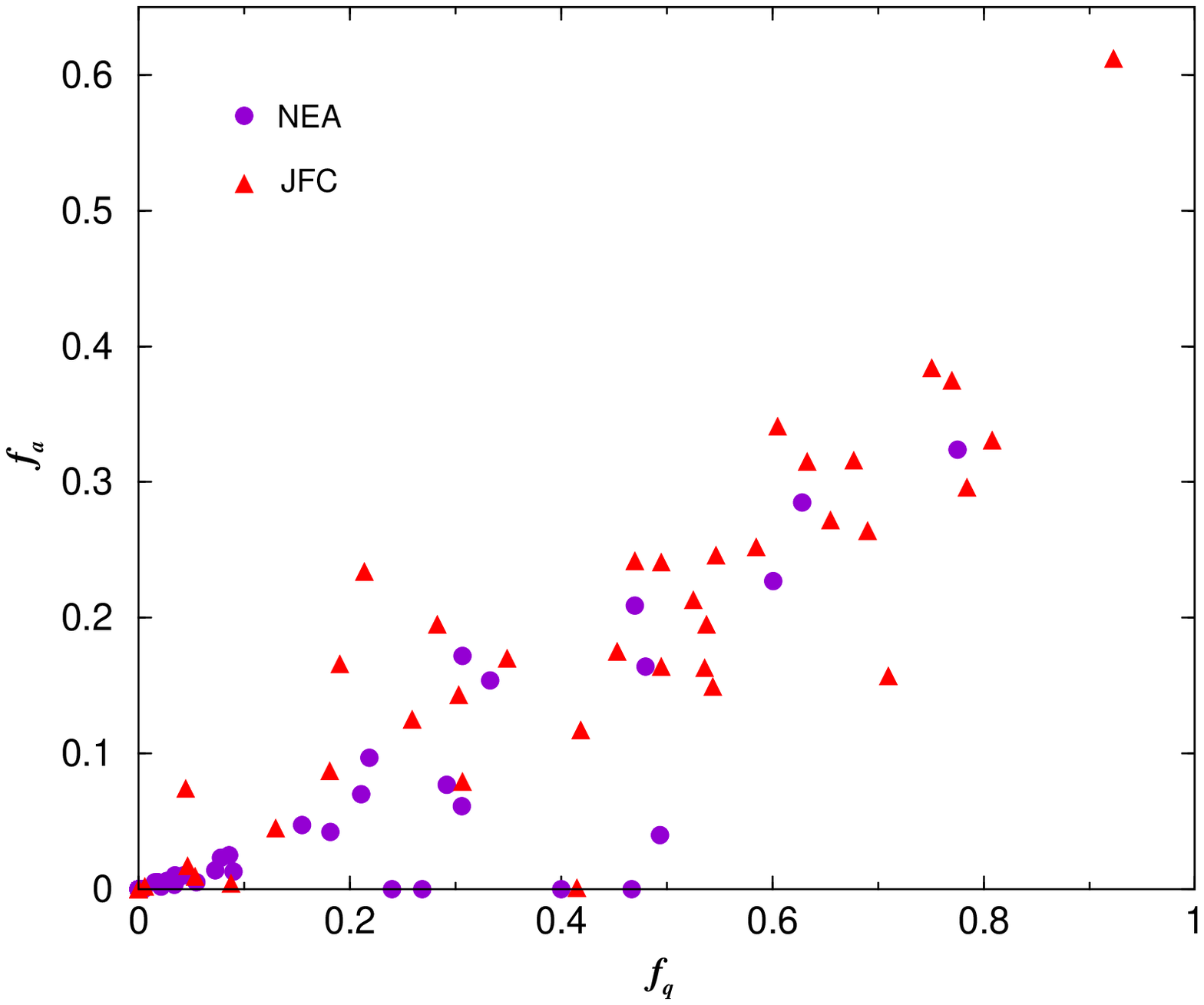}}
\caption{The indices $f_q$ versus $f_a$ for the JFCs and NEAs. We note in general a good correlation, except for a few NEAs in stable orbits, but with large periodic oscillations in their perihelion distances that drive them above 2.5 au (these are located on or near the $f_q$-axis and with $f_q > 0.2$).}
\label{}
\end{figure}

In Fig. 7 we plot the fractions $f_q$ versus $f_a$ for our samples of NEAs and JFCs. A large proportion of our studied NEAs and a few JFCs superpose on the origin ($f_q=f_a=0$). For the remainder, there is a good correlation between both parameters (we get correlation coefficients of 0.875 and 0.860 for the JFC and the NEA sample, respectively, in the latter excluding the four NEAs lying on the $f_q$-axis that move on stable orbits but with large oscillations of $q$).\\

\begin{figure}[h]
\resizebox{10cm}{!}{\includegraphics{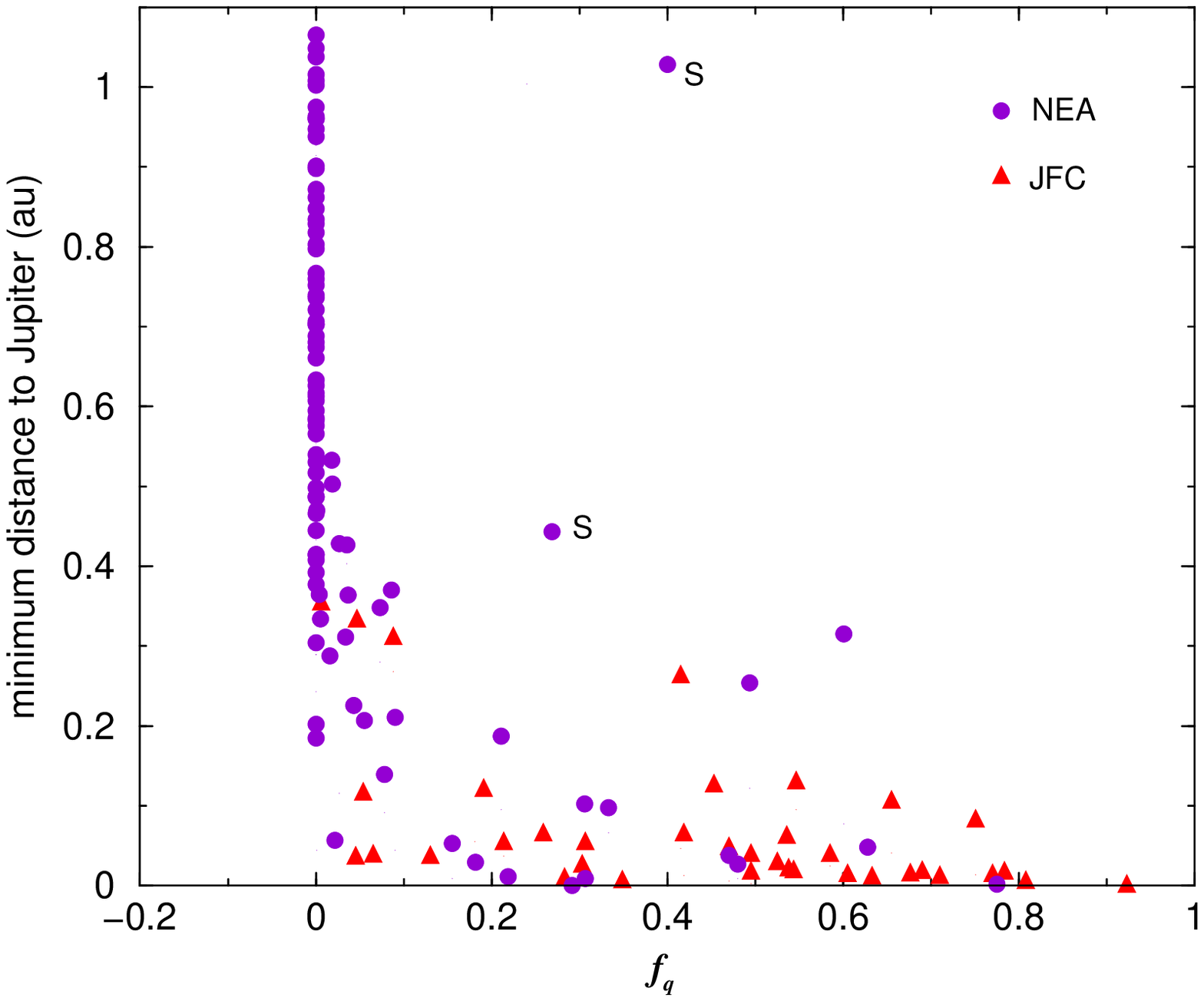}}
\caption{The index $f_q$ versus the distance of closest approach of each object (and its clones) to Jupiter. We can see that objects in unstable orbits (large values of $f_q$) experience very close approaches to Jupiter. These occurs for most of the JFCs and some NEAs. The letter "S" stands for those NEAs in stable orbits.}
\label{}
\end{figure}

In Fig. 8 we plot the distances of closest approach to Jupiter within 3 Hill radius, $d_{min}$, versus the fraction $f_q$ of the NEAs and JFCs of our samples (and its clones) during the past $10^4$ yr. We note that the computed $d_{min}$ values were only recorded for the original object, not for its clones. There are 45 NEAs and 1 JFCs that do not appear in the graph since they did not experience close encounters with Jupiter. As expected, when $f_q=0$, the objects tend to avoid close encounters with Jupiter at distances below 0.4 au (about one Hill radius). Most of the NEAs and some JFCs of our sample are in this situation. On the other hand, most objects experiencing encounters with Jupiter to distances $\lsim 1/3$ Hill radius ($\sim 0.15$ au) have $f_q > 0.2$. This is the case of the great majority of JFCs and a few NEAs.

\subsection{Dynamical transfer from the asteroid belt to NEA orbits}

There is a huge gap in the asteroid belt around the 2:1 MMR. Yet the gap is not completely empty. It contains a transient population of asteroids in the resonance injected from the adjacent populated regions by chaotic diffusion and the Yarkovsky effect \citep{Roig02}, or by collisional breaukup events \citep{Moon98}. We consider this transient population as the potential progenitors of the NEAs in the same resonance. To test this hypothesis, we studied the orbital evolution of a sample of observed asteroids evolving inside the 2:1 resonance with proper semimajor axes, $a_{p}$, in the range $3.27 < a_{p} < 3.28$ au, and proper eccentricities $e_{p} > 0.25$, taken from the database ASTDyS (\verb"hamilton.dm.unipi.it/astdys/"). We have 132 objects that fulfill these conditions whose perihelion distances fall in the range $1.7 \lsim q \lsim 3$ au, i.e. the most eccentric ones are close but still outside the NEA region.\\ 

The numerical integrations were carried out for 100 Myr with the code EVORB \citep{Fern02}, including the planets from Venus to Neptune, whereas Mercury was added to the Sun. No relativistic effects were taken into account. The initial osculant elements of the asteroids were taken from the ASTORB database (\verb"www.naic.edu/~nolan/astorb.html") for the epoch 2456600.5. The code EVORB works with a leapfrog scheme with a time step of 0.01 yr for this simulation, but when an encounter with a planet at less than 3 Hill radius takes place the integration switches to a Bulirsch-Stoer integrator, providing a good compromise bewteen precision and computing speed.\\ 

The numerical results show that the eccentricities of the asteroids tend to increase with time, so many of them become NEAs at a rate of $\sim 7$ per Myr, keeping all the way in the 2:1 MMR. It is interesting to note that if these bodies reach $q=1.3$ au in the 2:1 MMR, their aphelion distances will rise to $Q \sim 5.25$ au, so all of them will fulfill the condition $Q>4.8$ au imposed to our NEA sample. On the other hand, bodies reaching $q=1.3$ au in other resonances like 5:2 will barely attain $Q \sim 4.3$ au, so their perihelion distances will have to decrease even further to $q \lsim 0.8$ au in order to reach $Q \gsim 4.8$ au. This explains the overabundance of bodies in the 2:1 MMR in our sample.\\

The transfer rate of 7 objects Myr$^{-1}$ in the 2:1 MMR turns out to be several times greater than others that could be found in the literature \citep{Roig02}, most probably due to the rather high initial eccentricities of our chosen asteroids that place them near the exit door. More specifically, the direct source of a large fraction of our NEAs are what Roig et al. ({\it loc. cit.}) classified as the 'unstable' ones (near the boundary) and the Griquas of the 2:1 MMR, with typical lifetimes in the resonance of the order of a few 10 Myr to some 100 Myr respectively. These authors identified another group at the core of the resonance, called the Zhongguos, which seem to be stable over the solar system age, so they cannot be a source of NEAs. As shown in Fig. 9, there is a good match between the computed NEAs transferred from the asteroid belt and the observed ones in 2:1 MMR. A considerable fraction (19 of 132) of the computed population inside the 2:1 MMR ends up as sungrazers, and most of them keep inside or very close to the domain of the resonance at the moment of the collision with the Sun. This means that the high-eccentricity objects in the 2:1 MMR are not only an important source of NEAs but also of sungrazers.\\

The role of the 2:1 MMR in the transfer efficiency to NEA orbits is clearly seen when we compare the rate of asteroids that reach $q<1.3$ au starting inside the resonance, with the rate for nonresonant asteroids initially near the 2:1 resonance. To check this, we also integrated a sample of 132 asteroids with $e_{p} > 0.25$ but with proper semimajor axes $3.18 < a_{p} < 3.20$ au, namely at the left side of the resonance. From this sample we found a transfer rate of one per 10 Myr, that turns out to be approximately 70 times lower than the rate for bodies within the 2:1 resonance.\\

\begin{figure}[h]
\resizebox{10cm}{!}{\includegraphics{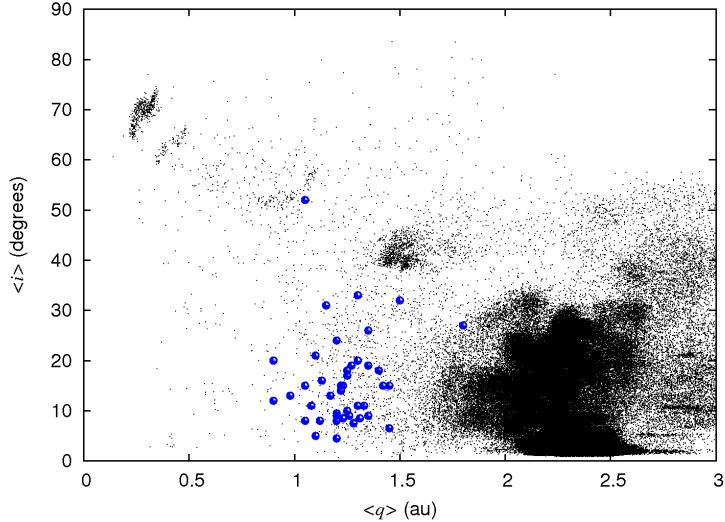}}
\caption{The evolution of a sample of asteroids in 2:1 MMR during 100 Myr (black dots) in the plane of mean inclinations versus mean perihelion distances. Every dot represents the mean value of $q$ and $i$ over a period of $2 \times 10^4$ yr, plotted at intervals of $2.5 \times 10^4$ yr. We also plot the locations of the NEAs in 2:1 MMR of our sample by considering their average $q$ and $i$ over the studied period of $2 \times 10^4$ yr (blue circles).}
\label{}
\end{figure}

We have also analyzed the efficiency of the 9:4 resonance in the transfer of bodies from the main belt to NEA orbits. We choose an identical number of 132 observed asteroids evolving inside the 9:4 MMR with proper semimajor axes $3.0287 < a_p < 3.0292$ au, and proper eccentricities $0.19 < e_p < 0.32$, which gives a distribution of initial perihelion distances in the range $1.95 \lsim q \lsim 2.65$ au, i.e. rather similar to the starting perihelion distances of the sample of 2:1 MMR bodies. We found a transfer rate of approximately 12 per 10 Myr, which turns out to be about 6 times lower than that found for the 2:1 resonance, but still more than 10 times greater than that derived for the nonresonant bodies close to the 2:1 MMR.

\subsection{Kozai mechanism}

The Kozai mechanism is the main responsible for the steady drop in $q$ observed in the numerical integrations. The dynamical evolution under this mechanism can be analyzed by means of the body's energy level curves, that depend on its semimajor axis and on the parameter $H=\sqrt{1-e^2} \cos i$. Kozai energy level curves guarantee a secular orbital evolution that keeps high perihelia (namely, low eccentricities) for asteroids inside or outside the 2:1 resonance as long as $H \gsim 0.85$, which is the case for most of the present population residing in the main asteroid belt. The diffusion in eccentricity that characterizes the resonance tends to lower the values of $H$. When it reaches values $H \lsim 0.7$, the topology of the energy level curves changes dramatically, making possible connections between high and very low values of $q$, correlated with oscillations in $\omega,i$ in a typical time interval of a few 1000 yr. The curves also depend on the resonant or nonresonant character of the orbital motion, in particular on the libration amplitude and center \citep{Gome05,Gall12}. These curves have a wider amplitude, connecting large with very low $q$ regions for nonresonant bodies, while they are flatter for resonant bodies of similar $H$ values. In other words, the Kozai effect is stronger for nonresonant bodies. This behavior is illustrated in Fig. 10 where we show the energy level curves for the 2:1 resonant case and the nonresonant case adjacent to the resonance. For example, a resonant asteroid with $H=0.6$ and $q=2$ au does not exhibit relevant variations in $q$ while captured in the resonance (Fig. 10, middle-left panel), but as soon as the resonant link is broken its perihelion can drop to 0.7 au (middle-right panel).\\

The tracks of four NEAs of our sample in the parametric plane ($\omega, q$) for the last $10^4$ yr are shown in Fig. 11. The panels on the left show two nonresonant objects, and the ones on the right two resonant objects inside the 2:1 resonance with Jupiter at  $a=3.28$ au. 2010 QE2 is evolving inside a typical secular Kozai mechanism at $a=3.35$ au with a small value of the parameter $H$, having an inclination that varies between 22 and 74 degrees correlated with strong perihelion variations. 2007 JF22 is also evolving inside a typical secular Kozai mechanism at $a=3.08$ au but with a high value of $H$, which constrains the range of possible values of $q$ within the rather narrow range $1 \lsim q \lsim 2$ au. 2006 AL8 is evolving inside the 2:1 resonance with a large libration amplitude and a low value of $H$ exhibiting a secular Kozai mechanism that generates large perihelion excursions and inclination variations between 35 and 66 degrees. Nevertheless we can see that the perihelion excursions of 2006 AL8 are smaller than those of the nonresonant 2010 QE2 body with a rather similar $H$ value. 
2002 VY94 is also inside the 2:1 resonance but with a larger value of $H$ that prevents it from reaching low perihelion values. We recall the cases of the NEAs of Fig. 2. The resonant object 2004 RU164 ($H=0.73$) shows a weak Kozai effect, whereas the nonresonant object 2004 QU24, with a similar $H=0.74$, exhibits a much stronger Kozai effect.

\begin{figure}[h]
\resizebox{10cm}{!}{\includegraphics{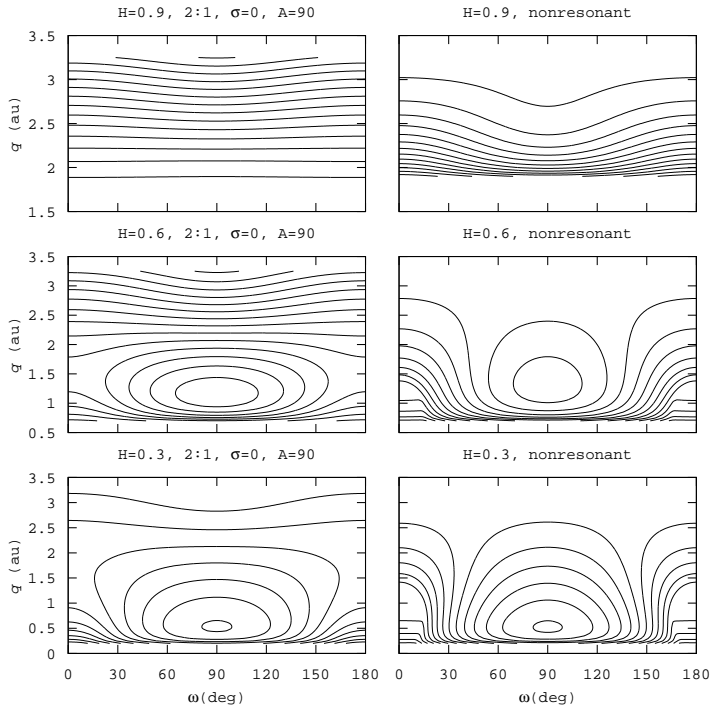}}
\caption{Kozai diagrams of fictitious bodies in the plane $q$ versus the argument of perihelion $\omega$ for three values of $H=0.9$, 0.6 and 0.3. The left column shows resonant cases with a libration center $\sigma = 0$ and amplitude $A = 90^{\circ}$, and the right column nonresonant ones.}
\label{}
\end{figure}

\begin{figure}[h]
\resizebox{10cm}{!}{\includegraphics{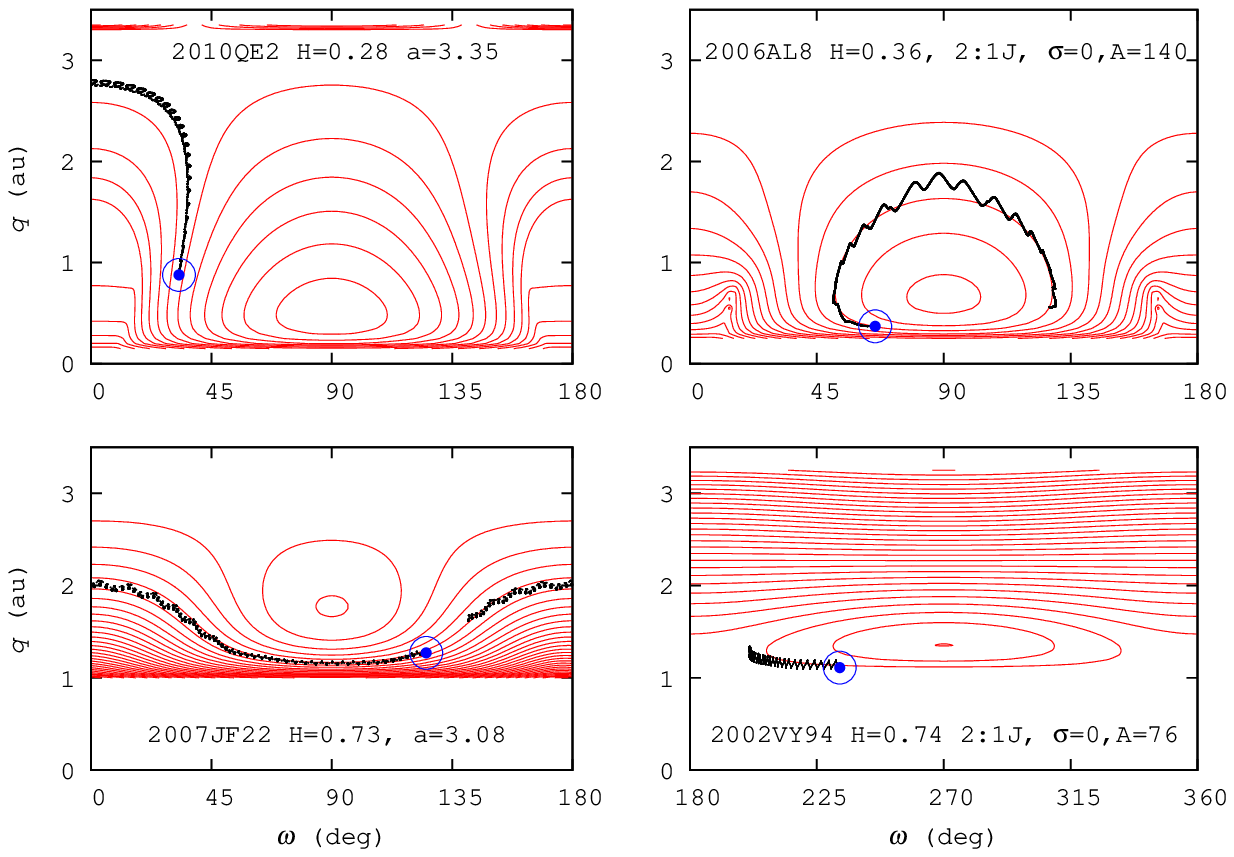}}
\caption{Kozai diagrams for the nonresonant NEAs 2010 QE2 and 2007 JF22, and the resonant ones 2006 AL8 and 2002 VY94. The circle indicates the position of the object at present. The track in bold indicates the motion of the object in this parametric plane in the last $10^4$ yr.}
\label{}
\end{figure}

\subsection{Sungrazing states}

Due to the action of the Kozai mechanism, several comets and asteroids will end up in sungrazing orbits \citep{Bail92}. Three objects of our sample of NEAs are currently with perihelion distances $q < 0.15$ au. These are: 2002 PD43, 2006 HY51 and 2008 HW1, and they have been in this state for the last 890 yr, 640 yr, and 300 yr, respectively. There is another object, 2003 EH1 that has now $q=1.189$ au but that reached a minimum $q=0.116$ au about 1500 yr ago, and stayed with $q < 0.15$ au between about 1750 and 1250 yr ago. This object is associated to the Quadrantid meteor shower \citep{Jenn04}. We show in Fig. 12 two of the sungrazers of our sample: 2002 PD43 and 2003 EH1. We also note that PD43 is in the 3:1 MMR with Jupiter.

\begin{figure}[h]
\begin{center}
\begin{tabular}{c c}
\resizebox{6cm}{!}{\includegraphics{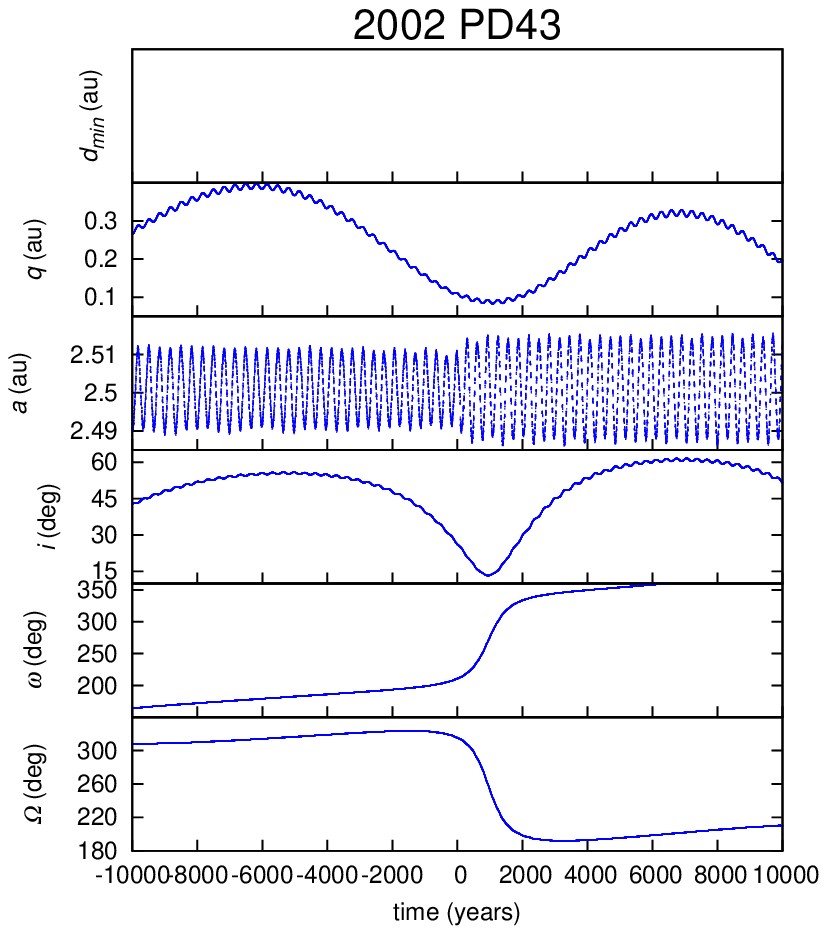}}
\label{}
&
\resizebox{6cm}{!}{\includegraphics{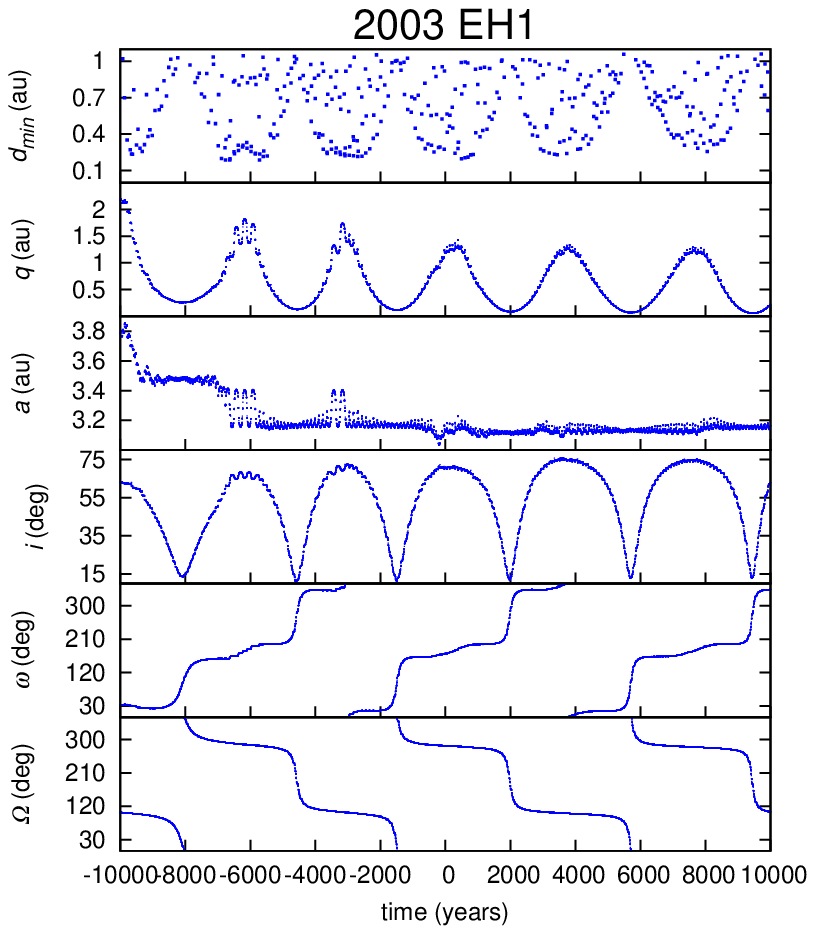}}
\label{}
\end{tabular}
\end{center}
\caption{Sungrazers. The NEA 2002 PD43 is currently with a $q<0.15$ au (left), whereas 2003 EH1 was sungrazer before (right). We note that PD43 does not experience encounters with Jupiter within three Hill's radius during the studied period.}
\end{figure}

\section{Discussion}

\subsection{Stable and unstable orbits}

Our results confirm that most NEAs move in stable orbits over the time scale of $2 \times 10^4$ yr of our computations, as shown by the parameters $f_q = 0$ and/or $f_a = 0$. On the other hand, most JFCs are found to move in unstable orbits, thus showing values for $f_q$, $f_a$ $>$ 0. We remind the reader that our conclusions were derived for bodies with $Q > 4.8$ au, namely that are potentially able to experience close encounters with Jupiter. Yet, a few NEAs are found to move in cometary orbits. These NEAs in Cometary Orbits (NEACOs) have unstable orbits with $f_q \gsim 0.25$ and $f_a \gsim 0.06$ and mix quite well with the JFCs in the plots of Figs. 5, 6 and 8. Table II brings the list of the most prominent NEACOs found in our sample. The times, $t_{NEA}$, elapsed since these bodies were transferred from a large-$q$ orbit ($>2.5$ au) to its current NEA orbit are relatively short, of the order of several hundreds to a couple thousands yr (i.e. $\sim 80-400$ revolutions). These short residence times in an orbit with $q < 1.3$ au are consistent with the physical lifetimes estimated for typical kilometer-size JFCs \citep{Fern02, Disi09}. We also have some other potential NEACOs (though of lower probability than the ones of Table II): 2010 LR68 and 2012 MA7, for which we find  $f_q \sim 0.2$.\\

\begin{table}[htb]
\centerline{Table II: NEACOs}
\begin{center}
\begin{tabular}{l l l l l l} \hline
Object & $H$ & $R^{(*)}$ (km) & $f_q$ & $f_a$ & $t_{NEA}$ (yr) \\ \hline
1997 SE5 & 14.8 & 3.25-2.30 & 0.628 & 0.285 & 365 \\
2000 DN1 & 19.8 & 0.33-0.23 & 0.307 & 0.172 & 850-1450 \\
2001 XQ & 19.2 & 0.43-0.30 & 0.306 & 0.061 & 1400 \\
2002 GJ8 & 19.4 & 0.39-0.28 & 0.292 & 0.077 & 670 \\ 
2002 RN38 & 16.9 & 1.24-0.87 & 0.470 & 0.209 & 485 \\ 
2003 CC11 & 19.1 & 0.45-0.32 & 0.775 & 0.324 & 400 \\ 
2003 WY25$^{(**)}$ & 20.9 & 0.20-0.14 & 0.480 & 0.164 & 600-2800 \\
2009 CR2 & 16.7 & 1.36-0.96 & 0.333 & 0.154 & 610 \\ 
2011 OL51 & 19.8 & 0.33-0.23 & 0.601 & 0.227 & 870-1640 \\ \hline
\end{tabular}
\end{center}
(*) The radii were computed by assuming geometric albedos $p_V=0.05-0.1$.\\
(**) It was identified with comet D/1819 W1 (Blanpain) and given the permanent name 289P/Blanpain.\\
\end{table}

\citet{Tanc14} has recently proposed a list of {\it Asteroids in Cometary Orbits} (ACOs), considering as a key criterion to define the cometary nature the {\it Minimum Orbital Intersection Distance} (MOID) parameter, namely the minimum distance between the orbits of the object and that of Jupiter. Tancredi identifies 331 ACOs from which 32 are in our NEA sample, including our 8 NEACOs of Table II. On the other hand, Tancredi includes 24 additional objects for which we find asteroid-like orbits, in most cases with $f_q = 0$ or close to zero. Object 2003 EH1 (Fig. 12, right) is a good example of an object included in Tancredi's ACO list that suffers many close encounters with Jupiter, though they are not able to break the regular pattern of its motion through most of the integration time. Therefore, even though the MOID parameter may be useful for a quick diagnostic, it is still necessary to perform direct integrations to know more accurately the degree of stability (or instability) of the orbit. 

\subsection{The physical nature of objects approaching the Earth}

Since stable orbits mean that objects may remain bound to small-$q$ orbits over time scales $>>$ $10^4$ yr, such objects will have to be able to withstand the intense solar radiation without significant erosion. On the other hand, active comets lose copious amounts of material in every passage, so they cannot survive for too long in the Earth's neighborhood as active bodies. Their orbits must accordingly be unstable, with short residence times in the near-Earth region before their discovery.\\ 

Among the NEOs, we can find different compositions that might be grouped into:\\

1) {\it Rocky bodies}: They are devoid of volatile material and their internal strength is rather high. They can withstand close passages by the Sun without losing measurable amounts of material, as is observed in many asteroids passing at heliocentric distances $< 0.25$ au \citep{Jewi13}. Collisions with meteoroids may be the only agent capable of triggering (dusty) activity.\\

2) {\it Rocky-acqueous bodies}: The matrix is built with the mineral (refractory) component, but they are carbon-rich and contain some water, either under the form of hydrated silicates, or even as water ice buried in their interiors. The mineral matrix provides a moderate internal strength, capable of withstanding close passages by the Sun with losses of material limited to the outer layers. These bodies may display some activity under the intense Sun's radiation, so they may become "rock comets" \citep{Jewi10}. Their progenitors may be the so called main-belt comets (MBCs) and their source region may be the outer asteroid belt where water molecules in the protoplanetary disk could have condensed into ice or bound to silicates. Spectroscopic observations of the MBCs, 133P/Elst-Pizarro and 176P/LINEAR, did not show the CN emission band at $\sim 3800$ {\AA}, setting an upper limit to the gas production rate $\sim 3$ orders of magnitude below that found for typical JFCs observed at similar heliocentric distances \citep{Lica11}. These observations show that whereas free sublimation of water ice is the main activation mechanism of comets near the Sun, other mechanisms that require little (or no) water sublimation are responsible for activating MBCs.\\

3) {\it Icy bodies}: The matrix is built as a very loose aggregate of ice and dust particles, so their bulk density and internal strength is very low \citep{Donn90, Sosa09}. They lose appreciable amounts of material in each passage by the Sun's vicinity by sublimation and frequent outbursts and breakups, so a typical one-km size nucleus has a very short physical lifetime. There has been a long discussion on whether they transit through stages of dormancy before final disintegration into meteoritic dust, or if they pass straight from an active nucleus to dust (cf. Fig. 1). These objects formed in the trans-Jovian region where they have remained until present in cold reservoirs: the trans-neptunian region and the Oort cloud.

\subsection{Thermally-induced activity of rocky-acqueous bodies close to the Sun}

The proximity to the Sun (distances $r \lsim 0.15$ au) can raise the surface temperature of a NEA with a visual geometric albedo $p_v \sim 0.05$ to about 1000 K. Expected physical effects are thermal fracture induced by strong temperature gradients, and dehydration in case the object contains minerals with bound water molecules, or OH radicals \citep{Jewi10}. These effects may be the source of some activity as, for instance, the release of water molecules as the chemically-bound water in hydrated silicates is progressively lost, starting at temperatures of about 600 K, and the ejection of dust particles that get enough energy to escape from the asteroid upon thermal fracture. \citet{Jewi13} searched for activity in 2002 PD43 with negative results, which might suggest that it is thermally resistant, perhaps with a composition rich in silicates like pyroxenes, olivines and metal, similar to the asteroids predominant in the inner belt of taxonomic classes S or M.\\ 

On the other hand, if 2003 EH1 has released material to give raise to the Quadrantid meteor stream, its composition might be different, perhaps rocky-acqueous as discussed before. Their different semimajor axes: 2.5 au for 2002 PD43, and 3.2 - 3.4 au for 2003 EH1, may give support to the idea of a different geochemical composition. If we assume that most of the mass of the Quadrantid was released when 2003 EH1 had a very small $q$ ($\lsim 0.15$ au), then the age of this meteor shower would be around 1500 yr. This is about three times the age estimated by \citet{Jenn04} based on the dispersion of the orbital parameters of the Quadrantid shower with respect to those of 2003 EH1. This discrepancy might be explained either because of an underestimation of the age of the breaukup that generated the Quadrantid, or because the breakup took place after the object reached the minimum $q$. For instance 500 yr ago 2003 EH1 had $q \sim 0.6$ au. As regards the physical nature of 2003 EH1, it has a typical asteroid orbit (cf. Fig. 12, right), so it may come from the outer asteroid belt and has a rocky-aqueous composition. Its activity might arise from thermal fracture and dehydration when the body's surface attains temperatures close to 1000 K \citep{Jewi10}.\\

(3200) Phaethon may be another example of an active asteroid whose activity is triggered by thermal fracture and/or decomposition of hydrated silicates upon close approach to the Sun ($q \lsim 0.14$ au). \citet{Li13} reported about one magnitude brightening at its 2009 and 2012 perihelion passages, and attributed it to the ejection of dust by the mechanism mentioned before. The visible and near infrared reflectance spectrum of this object is found to be similar to that produced by aqueously altered samples of CI/CM carbonaceous chondrites and hydrated silicates \citep{Lica07}. \citet{Dele10} found that (2) Pallas is the most likely parent body of Phaeton based on spectroscopic similarities between the latter and several members of the Pallas family, as well as numerical simulations that show a dynamical pathway through which fragments of Pallas can reach Phaethon-like orbits. 

\subsection{Taxonomic types and albedos among NEAs of our sample}

Unfortunately, the information available on these two physical parameters for our NEA sample is very scant. Only a small fraction of them have known albedos and/or taxonomic types. The collected data are shown in Table III, which have been drawn from \citet{Deme08}. Even though this is a small fraction of our total sample, it gives us a hint of the taxonomic types prevailing among NEAs with $Q > 4.8$ au. These are types C, D, and P, that correspond to low-albedo (around 0.05-0.06) primitive material, and linear spectra over visible wavelengths with neutral to red slopes. These taxonomic types prevail among the asteroids in the outer belt and the Jupiter's Trojans \citep{Deme08, Deme13}. These photometric features have also been associated with dead comets \citep{Jewi02, Lica08}. We also show in Table III the probabilities $P_{JFC}$, $P_{OB}$, for a given object to originate as a JFC, or in the outer asteroid belt, estimated by \citet{Deme08} following the dynamical model developed by \citet{Bott02}. In case $P_{JFC} + P_{OB} < 1$, the remaining probability, $1 - P_{JFC} + P_{OB}$, corresponds to an origin in the inner asteroid belt. Table III also includes our computed fractions $f_q$, $f_a$, and whether the body motion is locked in any particular resonance.\\ 

\begin{table}[htb]
\centerline{Table III: Taxonomic types and/or albedos of same NEAs}
\begin{center}
\begin{tabular}{l l l l l l l l} \hline
Object & Albedo$^{(*)}$ & Type$^{(*)}$ & $P_{JFC}^{(*)}$ & $P_{OB}^{(*)}$ & $f_q$ & $f_a$ & Res. \\ \hline
3552 Don Quixote & $0.045 \pm 0.003$ & D & 1.000 & 0.000 & 0.494 & 0.040 & 4:3 / 7:5 \\
1997 SE5 & - & T & 1.000 & 0.000 & 0.628 & 0.285 & - \\
1997 YM3 & - & C & 0.155 & 0.809 & 0.000 & 0.000 & 2:1 \\
1999 DB2 & - & S & 0.424 & 0.547 & 0.000 & 0.000 & - \\
1999 LT1 & - & C,F & 0.738 & 0.037 & 0.000 & 0.000 & 7:3 \\
2000 EB107 & - & D & 0.904 & 0.036 & 0.000 & 0.000 & 9:4 / 11:5 \\
2000 PG3 & $0.046 \pm 0.012$ & D & 0.929 & 0.025 & 0.000 & 0.000 & 5:2 \\
2000 WL10 & - & Xc & 0.926 & 0.041 & 0.004 & 0.000 & 2:1 \\
2001 RC12 & 0.080 & - & 0.077 & 0.849 & 0.000 & 0.000 & 2:1 \\
2001 XP1 & $0.03 \pm 0.01$ & P & 0.776 & 0.000 & 0.000 & 0.000 & - \\
2003 UL12 & 0.093 & - & 0.321 & 0.669 & 0.000 & 0.000 & 2:1 \\
2003 XM & - & T & 0.677 & 0.316 & 0.000 & 0.000 & 2:1 \\
2003 WY25$^{(**)}$ & - & D & 0.874 & 0.106 & 0.480 & 0.164 & - \\
2004 YZ23 & $0.05 \pm 0.02$ & P & 0.602 & 0.070 & 0.000 & 0.000 & - \\
2005 AB & $0.04 \pm 0.02$ & C & 0.268 & 0.712 & 0.000 & 0.000 & 2:1 \\
 \hline
\end{tabular}
\end{center}
(*) These data were taken from \citet{Deme08} and \citet{Fern05}.\\
(**) Now 289P/Blanpain.
\end{table}

Our results are partially in conflict with those obtained by \citet{Deme08} and \citet{Bott02} and tend to be much more pessimistic as regard the possible cometary nature of some NEAs. According to our dynamical criterion that defines a likely comet origin by the degree of instability of the objects's orbit (measured through the indices $f_q$ and $f_a$), we find a likely comet origin only for 1997 SE5 and 2003 WY25 (confirmed in the latter case by \citet{Jewi06} who found a weak coma around the object), in agreement with \citet{Deme08}'s results. Furthermore, we both agree on an asteroidal origin only for the cases of 1999 DB2, 2001 RC12 and 2003 UL12. On the other hand, we find a likely asteroidal origin for 3552 Don Quixote, 1997 YM3, 1999 LT1, 2000 EB107, 2000 PG3, 2000 WL10, 2001 XP1, 2003 XM, 2004 YZ23 and 2005 AB, while \citet{Bott02} and \citet{Deme08} find a cometary one. As we see in Table III, all these objects have $f_q = f_a = 0$ (except for the peculiar object 3552 whose situation was explained in Section 3.2). Furthermore, most of them move for all or most of the computed time in one of the main resonances with Jupiter, in particular the 2:1. These orbit characteristics suggest that they are stable over time scales $>> 10^4$ yr, typical of asteroids, not of comets. Two of these objects hop from one MMR to another one, while only two objects moving on stable orbits and the two in unstable orbits show a nonresonant motion.    

\subsection{Populations of NEAs and NEJFCs with $Q > 4.8$ au}

The discovery rate of NEAs shows a steep increase in the last decade thanks to the contribution from large sky surveys like LINEAR, Catalina, Siding Spring, NEAT, LONEOS and Spacewatch among others (Fig. 13). From the discovery rate we can make a rough estimation of the population, down to a limiting absolute magnitude $H=17.4$, that roughly corresponds to a radius $R \sim 1$ km for a typical geometric albedo 0.05. There are 13 observed NEAs with $Q > 4.8$ au and $H \leq 16$. Since the discovery rate of this sub-sample of brighter NEAs has dropped in the last decade or so (cf. Fig. 13), we can assume that it is near completion. To allow for some as yet undetected objects, we can set its total number at about 20. Furthermore, we can also assume that the cumulative luminosity function derived by \citet{Bott00} for the whole NEA population applies, namely

\begin{equation}
N(<H) = C \times 10^{\gamma(H-13)}
\end{equation}
where $N(<H)$ is the number of objects brighter than $H$, $C$ is a constant and $\gamma = 0.35 \pm 0.02$.\\

Under the previous assumptions we find

\begin{equation}
N(17.4) = N(16) \times 10^{\gamma(17.4-16)} = 62
\end{equation} 
There are 46 observed NEAs brighter than 17.4, which may represent $\sim 3/4$ of the total population. We may then conclude that even down to $H=17.4$ the NEA sample is close to completion.\\

\begin{figure}[h]
\resizebox{10cm}{!}{\includegraphics{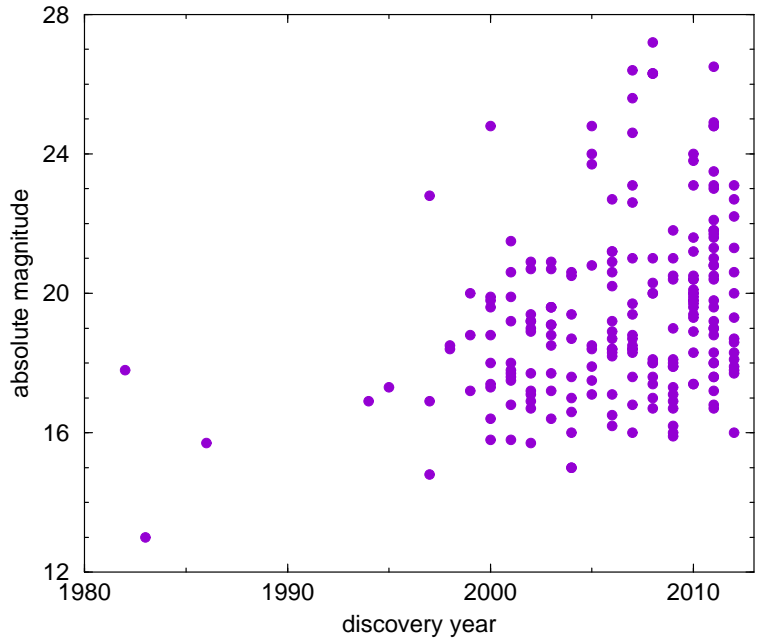}}
\caption{The discovery rate of NEAs with $Q > 4.8$ au.}
\label{}
\end{figure}


Among our sample of 139 NEAs with $Q > 4.8$ au we find 9 moving on typical cometary orbits, so we may argue that these bodies might be potential inactive comets. Yet 6 of the NEACOs are very faint: $H > 19$ ($R < 0.5$ km), so we may presume that they might be devolatized fragments of parent comets, rather than a comet that became dormant because of the buildup of an insulating dust mantle. One example of this might be the object 2003 WY25 (now re-defined as comet 289P/Blanpain) whose small size ($R \sim 0.2$ km) strongly suggests that it is a fragment of a former larger comet \citep{Jenn05}.\\

If we restrict ourselves to the sample of 46 observed NEAs with $H \leq 17.4$ ($R \gsim 1$ km), we find 3 NEACOs among them. Bearing in mind that this population may be $\sim 3/4$ of the total one of 62 (cf. eq.(5)), we might have an extra NEACO brighter than $17.4$ still to be discovered.\\

Let us now turn to the population of NEJFCs. \citet{Fern06} found a population of $24 \pm 10$ comets brighter than absolute total magnitude $H_{10} = 10$, which roughly corresponds to a nuclear magnitude $H_N \sim 17.6$, or a nucleus radius $R_N \sim 1$ km for a geometric albedo $p_v = 0.04$. This result is consistent with the one derived by \citet{Disi09} who found a population of $25 \pm 5$ JFCs with $H_N < 17.6$ and $q < 1.5$ au.\\

If we assume that the NEACOs of our sample are indeed of cometary nature, the ratio of inactive to active comets for radii $R \geq 1$ km will be 

\begin{equation}
\frac{n_{inactive}}{n_{active}} \simeq \frac{4}{24} \simeq 0.17
\end{equation}

This ratio may be taken as an upper limit because the NEACOs might well be bona fide asteroids that attained unstable orbits and are in the process of being ejected. Therefore, it is possible that the ratio of dormant/defunct comets is zero, at least for the larger bodies of our sample with radii $R > 1$ km. The situation may be different for the smaller objects, since devolatized fragments of disintegrated comets might fall in this population, as might be the case of 289P (2003 WY25).\\

The result of eq.(6) only concerns objects with $Q > 4.8$ au, which are the ones more prone to have unstable, comet-like orbits due to their potential close approaches to Jupiter. We expect that for smaller aphelion distances the orbits are more stable as the objects cannot get too close to Jupiter, so the result of eq.(6) should definitely be taken as an upper limit.

\section{Conclusions}

We summarize below the main results of our work. We stress that they have been derived for objects approaching or crossing Jupiter's orbit (aphelion distances $Q > 4.8$ au). Our main conclusions are:

\begin{enumerate}

\item As expected, most NEAs are found to move on highly stable 'asteroidal' orbits, whereas most JFCs move on unstable 'cometary' orbits during the studied period of $\pm 10^4$ yr, centered on the present.

\item Despite the previous conclusion, we detected a small group of NEAs that move on typical cometary orbits (that we called NEACOs). The objects of our sample that exhibit the most unstable, comet-like orbits are: 1997 SE5, 2000 DN1, 2001 XQ, 2002 GJ8, 2002 RN38, 2003 CC11, 2009 CR2 and 2011 OL51. Object 2003 WY25 also included in this group, has been identified with comet 289P/Blanpain.

\item Under the most optimistic assumption that all the NEACOs in our sample are of comet origin, we find a ratio of inactive to active comets of $\sim 0.17$, but it could be even smaller if these NEACOs turn out to be {\it bona fide} asteroids. Such a small ratio ($\lsim 0.17$) suggests that the periods of dormancy of comets in the inner planetary region are very short, and could be even absent. More likely, comets continue active until disintegration, possibly leaving behind in some cases large fragments of devolatized material that might enlarge the population of faint NEAs (sizes smaller than a few hundred meter). Object 2003 WY25 might be one of these fragments.

\item In particular, our results do not support previous claims that the bodies 3552 Don Quixote, 1997 YM3, 1999 LT1, 2000 EB107, 2000 PG3, 2000 WL10, 2001 XP1, 2003 XM, 2004 YZ23 and 2005 AB are extinct or dormant comets. We find that these bodies have moved on very stable orbits for at least the past $10^4$ yr, which would imply physical lifetimes much longer that the ones estimated for small (km-size) comets.

\item The locations of the semimajor axes of the NEAs in stable orbits are found to concentrate around the values corresponding to some of the main mean motion resonances with Jupiter. More than 40\% of the NEAs with $Q > 4.8$ au are in the 2:1 MMR, the immediate sources can be identified with the more chaotic main-belt asteroids near the boundary of this resonance, and the Griquas (that are marginally unstable bodies in the same resonance).

\item The transfer of bodies from the asteroid belt to NEA orbits is also partially ruled by the Kozai mechanism that produces a coupling between the inclination and the perihelion distance of the body, in such a way that the minimum of $i$ corresponds to the minimum of $q$ and vice versa. The Kozai mechanism can thus force the bodies to reach high inclinations, higher than those reached by most JFCs and direct some objects to sungrazing orbits.

\item Three objects of our sample of NEAs are currently with perihelion distances $q < 0.15$ au. These are: 2002 PD43, 2006 HY51 and 2008 HW1, and they have been in this state for the last 890 yr, 640 yr, and 300 yr, respectively. There is another object, 2003 EH1 that has now $q=1.189$ au but that reached a minimum $q=0.116$ au about 1500 yr ago, and stayed with $q < 0.15$ au between about 1750 and 1250 yr ago. It could have attained temperatures high enough to produce thermal cracks on its surface that could have led to the release of gas and dust if it has a rocky-acqueous composition. As mentioned, this object is associated to the Quadrantid meteor shower.

\end{enumerate}

\bigskip

\noindent{\bf Acknowledgments}

\medskip

This research was partially supported by the project FCE\_2\_2011\_1\_6990 of the Agencia Nacional de Investigaci\'on e Innovaci\'on (ANII). We want to thank Gabriel Santoro for his collaboration with the artwork. We also want to thank Ramon Brasser who, as a referee, made important comments and criticisms on the text that helped to clarify some points and improve the general presentation of the results.

\end{document}